\documentclass[prd,nofootinbib,preprint,superscriptaddress]{revtex4}
\usepackage{amsmath, amssymb, amsthm, graphicx, epsfig, fancyhdr,epsfig, slashed, mathrsfs}
\usepackage{caption}
\usepackage{braket}
\usepackage{tikzsymbols}
\usepackage{tikz}
\usepackage{pgffor}
\usepackage{tikz-feynman}
\tikzfeynmanset{compat=1.1.0}
\usepackage{natbib}
\usepackage{float}
\usetikzlibrary{shadows}
\usepackage{pifont}
\usepackage{booktabs}
\usetikzlibrary{arrows.meta, positioning, shapes, decorations.markings}
\usepackage{adjustbox}
\usepackage{caption}
\usepackage{tikz,xcolor,hyperref}

\definecolor{lime}{HTML}{A6CE39}
\DeclareRobustCommand{\orcidicon}{
	\begin{tikzpicture}
	\draw[lime, fill=lime] (0,0) 
	circle [radius=0.2] 
	node[white] {{\fontfamily{qag}\selectfont \tiny ID}};
	\draw[white, fill=white] (-0.0625,0.095) 
	circle [radius=0.007];
	\end{tikzpicture}
	\hspace{-2mm}
}

\foreach \x in {A, ..., Z}{\expandafter\xdef\csname orcid\x\endcsname{\noexpand\href{https://orcid.org/\csname orcidauthor\x\endcsname}
			{\noexpand\orcidicon}}
}

\begin{document}

\title{Three-Flavoured Thermal Leptogenesis Without Tuning}
\author{Angus Spalding \orcidC{}}
\email{angus.spalding1@gmail.com}
\affiliation{School of Physics and Astronomy, University of Southampton,
Southampton SO17 1BJ, United Kingdom}

\begin{abstract}
\noindent
We investigate thermal leptogenesis in an extension of the Type-I seesaw in which a real singlet scalar couples non-diagonally to the right-handed neutrinos. The interactions $-y_{ij}\phi N_iN_j$, with $i\neq j$, introduce new sources of CP violation in heavy-neutrino decays with a negligible enhancement of the standard leptonic washout. We derive an analytic upper bound on the scalar-induced CP asymmetry and present numerical results using fully flavoured Boltzmann equations and the density-matrix formalism. We show that the observed baryon asymmetry can be generated with a hierarchical right-handed-neutrino spectrum at the electroweak scale. In the minimal Type-I seesaw, lowering the scale of decay leptogenesis typically requires either a tuned quasi-degeneracy of the right-handed-neutrino masses or cancellations between the tree-level and one-loop contributions to the light-neutrino masses, while hierarchical leptogenesis instead requires a high scale that aggravates the Higgs hierarchy problem. This singlet-scalar extension appears to avoid these tensions, allowing successful leptogenesis with a hierarchical right-handed-neutrino spectrum down to the electroweak scale.
\end{abstract}

\maketitle

\tableofcontents
\section{Introduction}
Neutrino oscillation experiments \cite{Fukuda_2001, Fukuda_2002, Ahmad_2002, Abe_2016, Smirnov_2016, Aartsen_2018} establish that neutrinos are massive, requiring an extension of the Standard Model. A minimal realization is the Type-I seesaw mechanism \cite{Minkowski:1977sc, Gell-Mann:1979vob, Yanagida:1979as, Mohapatra:1980yp}, where heavy right-handed neutrinos generate suppressed light-neutrino masses through Yukawa interactions with the Higgs field. The heavy right-handed neutrinos of the Type-I seesaw can also account for the baryon asymmetry of the Universe through leptogenesis \cite{Fukugita:1986hr}. Out-of-equilibrium, $CP$-violating decays of heavy right-handed neutrinos generate a lepton asymmetry, which is partially converted into a baryon asymmetry by electroweak sphalerons \cite{Luty:1992un, Giudice_2004, Covi_1996, Buchm_ller_2005, Sakharov:1967dj, Kolb:1979qa, Khlebnikov:1988sr}.\\
Leptogenesis is most commonly studied in the context of thermal production of right-handed neutrinos, where these states are generated through scattering processes in the early Universe after reheating, assuming a radiation-dominated Universe throughout \cite{Buchm_ller_2005, McKenna:2026xar, Giudice_2004}.\\
A crucial ingredient often neglected in leptogenesis analyses is flavour. Below temperatures of approximately $10^{12}\,\mathrm{GeV}$, charged-lepton Yukawa interactions progressively become rapid enough to resolve individual lepton flavours. As a result, each flavour experiences different washout and CP-violating effects, altering the final baryon asymmetry. A consistent treatment therefore requires flavour-resolved Boltzmann equations or, for certain regions of parameter space, the full set of density-matrix equations to capture flavour decoherence and oscillations \cite{Nardi_2006, Nardi:2006fx, Abada_2006, Antusch_2006, Blanchet_2007, DeSimone:2006nrs, Cirigliano_2010, Simone_2007, Racker_2012, Moffat_2018, Ulysses, Ulysses2, blanchet2013leptogenesisheavyneutrinoflavours}.\\
Minimal Type-I seesaw leptogenesis also faces a well-known tension between the high scale required for successful thermal leptogenesis and electroweak naturalness. In the hierarchical regime, successful leptogenesis typically requires $M_1 \gtrsim \mathcal{O}(10^9)\,\mathrm{GeV}$, while such heavy right-handed neutrinos can induce unnaturally large radiative corrections to the Higgs mass parameter \cite{Vissani:1997ys, Clarke:2015gwa, Moffat_2018}. Two commonly explored ways of alleviating this tension are resonant leptogenesis and flavour effects. In resonant leptogenesis, the CP asymmetry is enhanced by taking the RHN masses to be nearly degenerate \cite{Klaric_2021, Pilaftsis_2004, Riotto_2007, Garbrecht_2014, Garny_2013, Anisimov_2006, Spalding:2026jia, Ghoshal:2025iil}; however, this requires a significant tuning of the heavy-neutrino mass spectrum. Alternatively, flavour effects can lower the viable leptogenesis scale \cite{Moffat_2018}, although sufficiently low-scale solutions involve cancellations between the tree-level and one-loop contributions to the active-neutrino masses. These considerations motivate extensions of the minimal Type-I seesaw in which successful leptogenesis can be achieved at lower scales without relying on either near-degenerate RHN masses or large cancellations between the tree-level and one-loop contributions to the light-neutrino masses.\\
A simple extension is to introduce a scalar singlet to the Type-I seesaw. This scalar is allowed by symmetries to couple to the right-handed neutrinos through Yukawa interactions of the form
$-y_{ij}\phi\overline{N_i^c}N_j$. Allowing the scalar Yukawa matrix to be non-diagonal in the heavy-neutrino mass basis gives rise to additional CP-violating contributions to the decays of the heavier right-handed neutrinos. These contributions can enhance the generated asymmetry without a
corresponding enhancement of the standard leptonic washout. Leptogenesis in scalar-extended Type-I seesaw models has been studied previously in
Refs.~\cite{LeDall:2014too,Alanne:2018brf}. The purpose of the present work is
not to introduce this model as a new mechanism, but to provide a more complete
treatment of its flavour dynamics. In particular, we derive an analytic upper
bound on the scalar-induced CP asymmetry and study the generation of the baryon
asymmetry including full lepton-flavour effects.\\
In this paper we present the full flavoured calculation and show that the scale of leptogenesis can be lowered to the electroweak scale without invoking the tuning concerns of minimal Type-I leptogenesis.\\
\textit{This paper is organised as follows:} Section \ref{sec:model} reviews our model Lagrangian and the new CP-asymmetry terms. In Section \ref{sec:flavourless} we consider the one-flavoured regime. Section \ref{sec:flavourful} presents the fully flavoured analysis, while Section \ref{sec:DME} treats the intermediate regime using the density-matrix equations. Finally, we conclude in Section \ref{sec:conclusion}.

\section{Model}
\label{sec:model}

In this section we introduce the scalar-extended Type-I seesaw model considered
throughout this work. We first define the particle content and relevant
interactions before deriving the scalar-induced contribution to the CP
asymmetry and obtaining an analytic upper bound on this new contribution.

\subsection{The Model}

We extend the Type-I seesaw by introducing a real gauge-singlet scalar $\phi$,
coupled to the right-handed neutrinos through a generally non-diagonal Yukawa
matrix $y_{ij}$. The relevant Lagrangian terms are
\begin{equation}
    \mathcal{L}_{\text{model}}
    \supset
    -Y_{\alpha i}\,\overline{L^\alpha_L}\,\tilde{H}\,N_i
    -\frac{1}{2}M_i\,\overline{N_i^C}N_i
    -\frac{1}{2}y_{ij}\phi\,\overline{N_i^C}N_j
    +\text{h.c.}
\end{equation}
where $L_\alpha$ denotes the lepton doublet of flavour
$\alpha=e,\mu,\tau$, $\tilde{H}=i\sigma_2H^*$ is the conjugate Higgs field, and $M_i$ are the heavy Majorana masses. After electroweak symmetry breaking, the Higgs acquires a vacuum expectation value $v= 174~\text{GeV},$ generating the Dirac mass matrix $m_D=vY$. In the limit $M_i\gg m_D$, the seesaw relation is obtained. The general solution for the Yukawa matrix $Y$ can be written as\cite{Casas_2001}
\begin{equation}
    Y = \frac{1}{v}\,
    U\,\sqrt{m_\nu}\,R\,\sqrt{M},
    \label{eq:CI}
\end{equation}
where $U$ is the PMNS matrix, $m_\nu$ the diagonal light-neutrino mass
matrix, $M$ the heavy-neutrino mass matrix, and $R$ is a complex orthogonal
matrix. This parametrisation, known as the Casas--Ibarra form, ensures that
the seesaw relation is automatically satisfied. The decay rate of a right-handed neutrino can then be written in the compact
form
\begin{equation}
    \Gamma_{N_i}
    =
    \frac{(Y^\dagger Y)_{ii}}{8\pi}\,M_i
    =
    \frac{M_i^2\tilde m_i}{8\pi v^2}\,.
\end{equation}
This defines the effective neutrino mass
\begin{equation}
    \tilde m_i
    \equiv
    \sum_j m_j |R_{ji}|^2\,.
\end{equation}
The effective masses $\widetilde m_i$ therefore provide a convenient measure
of the strength of the standard right-handed-neutrino interactions. 

\subsection{CP Asymmetry}

\begin{figure}[h!]
\captionsetup{justification=centering}
\centering
\begin{tikzpicture}

  \coordinate (Ni1) at (0,0);
  \coordinate (v1) at (1.5,0);
  \coordinate (ell1) at (3,1);
  \coordinate (phi1) at (3,-1);
  \draw[thick] (Ni1) -- (v1) node[midway, above] {\( N_i \)};
  \draw[thick] (v1) -- (ell1) node[midway, above] {\( \ell \)};
  \draw[dashed] (v1) -- (phi1) node[midway, below] {\( H^\dagger \)};
  \node at (1.5, -1.4) {(a)};

  \begin{scope}[xshift=4cm]
    \coordinate (Ni2) at (0,0);
    \coordinate (loopL) at (1.5,0);
    \coordinate (loopR) at (3,0);
    \coordinate (v2) at (4.5,0);
    \coordinate (ell2) at (6,1);
    \coordinate (phi2) at (6,-1);
    \draw[thick] (Ni2) -- (loopL) node[midway, above] {\( N_i \)};
    \draw[thick] (loopR) -- (v2) node[midway, above] {\( N_j \)};
    \draw[thick] (v2) -- (ell2) node[midway, above] {\( \ell \)};
    \draw[dashed] (v2) -- (phi2) node[midway, below] {\( H^\dagger \)};
    \draw[dashed] (loopL)
      arc[start angle=180, end angle=0,
          x radius=0.75cm, y radius=0.75cm]
      node[midway, above] {\( H \)};
    \draw[thick] (loopL)
      arc[start angle=-180, end angle=0,
          x radius=0.75cm, y radius=0.75cm]
      node[midway, below] {\( \ell \)};
    \node at (3, -1.4) {(b)};
  \end{scope}

  \begin{scope}[xshift=11cm]
    \coordinate (Ni3) at (0,0);
    \coordinate (vtx) at (1.5,0);
    \coordinate (Nj) at (3,0);
    \coordinate (ellOut) at (4.5,1);
    \coordinate (phiOut) at (4.5,-1);
    \coordinate (ellLoop) at (3,-1);
    \coordinate (phiLoop) at (3,1);
    \draw[thick] (Ni3) -- (vtx) node[midway, above] {\( N_i \)};
    \draw[thick] (vtx) -- (ellLoop) node[midway, below] {\( \ell \)};
    \draw[dashed] (vtx) -- (phiLoop) node[midway, above] {\( H \)};
    \draw[thick] (ellLoop) -- (phiLoop)
      node[midway, right] {\( N_j \)};
    \draw[thick] (phiLoop) -- (ellOut)
      node[midway, above right] {\( \ell \)};
    \draw[dashed] (ellLoop) -- (phiOut)
      node[midway, below right] {\( H^\dagger \)};
    \node at (2.3, -1.4) {(c)};
  \end{scope}

\end{tikzpicture}

\caption{\it Feynman diagrams contributing to the CP asymmetry:
(a) tree-level, (b) self-energy, and (c) vertex diagrams.}
\label{fig:cpasymmetrydiagrams}
\end{figure}
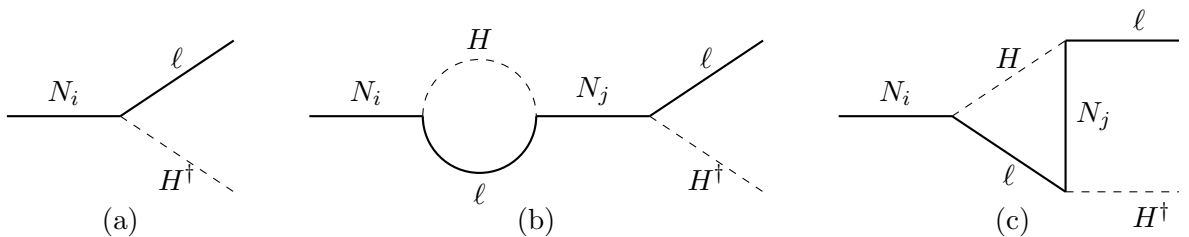

The lepton asymmetry is generated by the interference between the tree-level
decay $N_i\rightarrow L_\alpha H$ and its one-loop corrections. The relevant
Feynman diagrams for standard Type-I leptogenesis are shown in
Figure~\ref{fig:cpasymmetrydiagrams}.

The size of the asymmetry generated in the decay of each heavy neutrino $N_i$
to leptons of flavour $\alpha$ is quantified by the CP-asymmetry parameter
$\epsilon^i_\alpha$, defined as
\cite{Buchm_ller_2005, Di_Bari_2012}
\begin{equation}
    \epsilon^i_\alpha
    =
    \frac{
        \Gamma(N_i\rightarrow L_\alpha H)
        -
        \Gamma(N_i\rightarrow \bar{L}_\alpha H^\dagger)
    }{
        \displaystyle
        \sum_\beta
        \left[
        \Gamma(N_i\rightarrow L_\beta H)
        +
        \Gamma(N_i\rightarrow \bar{L}_\beta H^\dagger)
        \right]
    }\,.
\end{equation}
The total CP asymmetry is obtained by summing over charged-lepton flavours,
\begin{equation}
    \epsilon_i = \sum_\alpha \epsilon^i_\alpha\,.
\end{equation}
Evaluating the Feynman diagrams in the Type-I seesaw gives the standard result
\cite{blanchet2013leptogenesisheavyneutrinoflavours, Covi_1996}
\begin{equation}
    \epsilon^i_\alpha
    =
    \frac{1}{8\pi(Y^\dagger Y)_{ii}}
    \sum_{j\neq i}
    \left[
    \operatorname{Im}
    \left(
        Y^*_{\alpha i}Y_{\alpha j}(Y^\dagger Y)_{ij}
    \right)
    f_2\left(\frac{x_j}{x_i}\right)
    +
    f_1\left(\frac{x_j}{x_i}\right)
    \operatorname{Im}
    \left(
        Y^*_{\alpha i}Y_{\alpha j}(Y^\dagger Y)_{ji}
    \right)
    \right].
\end{equation}
where
\begin{equation}
    x_i = \frac{M_i^2}{M_1^2},
    \qquad
    f_1(x) = \frac{1}{x-1},
    \qquad
    f_2(x)
    =
    \sqrt{x}
    \left[
        (1+x)\ln\left(\frac{1+x}{x}\right)
        -
        \frac{2-x}{1-x}
    \right].
\end{equation}
This perturbative expression applies away from the resonant regime, where
finite-width effects associated with nearly degenerate heavy-neutrino masses
become important.

\begin{figure}[h!]
\centering
\begin{tikzpicture}
  \begin{feynman}
    \vertex (a) at (-6,0);
    \vertex (b) at (-4,0);
    \vertex (c) at (-2,0);
    \vertex (d) at (0,0);
    \vertex (e) at (2,1);
    \vertex (f) at (2,-1);

    \diagram* {
      (a) -- [plain, edge label=\(N_i\)] (b),

      (b) -- [half left, scalar, looseness=1.3,
              edge label=\(\phi\)] (c),
      (b) -- [half right, plain, looseness=1.3,
              edge label=\(N_j\)] (c),

      (c) -- [plain, edge label=\(N_k\)] (d),

      (d) -- [fermion, edge label=\(\ell\)] (e),
      (d) -- [scalar, edge label'=\(H\)] (f),
    };
  \end{feynman}
\end{tikzpicture}

\caption{\it Scalar-induced one-loop contribution to the CP asymmetry,
with $j<i$ for an absorptive contribution and $k\neq i$.}
\label{fig:newCPdiagram}
\end{figure}
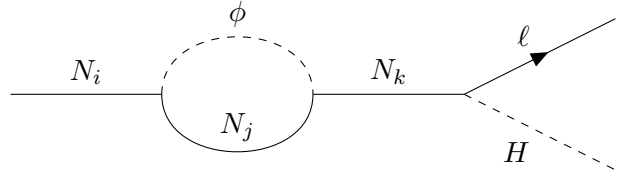

The off-diagonal scalar couplings introduce an additional one-loop
contribution to $N_i$ decay, shown in Figure~\ref{fig:newCPdiagram}.
Its interference with the tree-level amplitude provides a new source of
CP violation.

The relevant Yukawa structure is derived explicitly in
Appendix~\ref{sec:appendix}. The resulting scalar-induced contribution to the
flavoured CP asymmetry is
\begin{equation}
    \epsilon^\phi_{i\alpha}
    =
    \frac{1}{4\pi(Y^\dagger Y)_{ii}}
    \sum_{\substack{j<i\\k\neq i}}
    \operatorname{Im}
    \left[
        Y^*_{\alpha i}Y_{\alpha k}
        y_{ij}y_{jk}
    \right]
    \frac{M_i+M_k}{M_i^2-M_k^2}
    f(M_i^2,M_j^2,M_\phi^2)\,.
\end{equation}
The corresponding unflavoured scalar contribution is obtained by summing over
charged-lepton flavours,
\begin{equation}
    \epsilon_i^\phi
    =
    \sum_\alpha \epsilon^\phi_{i\alpha}\,.
\end{equation}
The loop function is
\begin{equation}
    f(x,y,z)
    =
    \frac{\sqrt{\lambda(x,y,z)}}{2x}
    \left(
        \frac{x+y-z}{2\sqrt{x}}
        +
        \sqrt{y}
    \right),
\end{equation}
where
\begin{equation}
    \lambda(x,y,z)
    =
    x^2+y^2+z^2-2xy-2xz-2yz
\end{equation}
is the K\"all\'en function. A non-zero absorptive
contribution requires
\begin{equation}
    M_i > M_j + M_\phi\,.
\end{equation}
For the ordered spectrum $M_1<M_2<M_3$, the scalar-induced contribution
therefore occurs only in the decays of $N_2$ and $N_3$, with the corresponding
scalar-mediated channel required to be kinematically open. In particular, the
new off-diagonal scalar interactions do not generate a corresponding
contribution to $\epsilon_1$. We therefore focus primarily on $N_2$-dominated leptogenesis where the lightest right-handed neutrino is weakly coupled to the standard model. For the $N_2$-dominated case, and in the hierarchical limit
$M_3\gg M_2\gg M_1,M_\phi$, Appendix~\ref{sec:appendix} gives the analytic upper bound
\begin{equation}
    |\epsilon^\phi_2|_{\max}
    =
    \frac{|y_{21}y_{13}|}{16\pi}
    \sqrt{\frac{M_2}{M_3}}
    \sqrt{\frac{\tilde m_3}{\tilde m_2}}
    \,.
\end{equation}
The corresponding bound on the asymmetry generated in an individual
charged-lepton flavour is
\begin{equation}
    |\epsilon_{2\alpha}^{\phi}|_{\max}
    =
    \frac{|y_{21}y_{13}|}{16\pi}
    \sqrt{\frac{M_2}{M_3}}
    \frac{\sqrt{\tilde m_{2\alpha}\tilde m_{3\alpha}}}
         {\tilde m_2}
    \,.
\end{equation}
Here $\tilde m_{i\alpha}/\tilde m_i$ gives the tree-level flavour projection
of $N_i$ onto flavour $\alpha$. The individual flavoured bounds are therefore
correlated and cannot in general be saturated independently. The bound shows that the scalar-induced CP asymmetry is controlled by the new couplings $y_{21}y_{13}$, while its dependence on the neutrino Yukawa sector enters through the effective neutrino masses. This separation allows the CP-violating source to be enhanced without a corresponding enhancement of the standard leptonic washout.

\section{Unflavoured Leptogenesis}
\label{sec:flavourless}

In this section we consider the simplest regime, the single-flavour limit, in
which the generated asymmetry and washout are effectively described by a
single flavour direction.  To compute the baryon asymmetry for a specific model, it is necessary to solve
the relevant Boltzmann equations, which track the evolution of the comoving
abundances of the scalar field $N_\phi$, the right-handed neutrinos (RHNs)
$N_{N_i}$, and the total $B-L$ asymmetry $N_{B-L}$. Throughout this work, we
normalise number densities following the convention adopted in
\cite{Ulysses, Ulysses2, Buchm_ller_2005, Granelli:2026goh} by dividing by
\begin{equation}
    N_{\rm norm} = \frac{8}{3\pi^2}T^3.
\end{equation}
This makes the equations numerically easier to solve by evolving dimensionless variables.

\subsection{Boltzmann Equations}
For the general three-right-handed-neutrino system, the evolution of the
heavy-neutrino abundances, the scalar abundance, and the total $B-L$
asymmetry is governed by \footnote{If the sole objective of the calculation is to determine the baryon asymmetry, the Boltzmann equation for the scalar may be omitted.}

\begin{align}
\frac{dN_{\phi}}{dz}
&=
\sum_{i>j}
D^\phi_{ij}\,
\Delta_{ij}^{\phi},
\\[2mm]
\frac{dN_{N_i}}{dz}
&=
-D_i\left(N_{N_i}-N_{N_i}^{\rm eq}\right)
+\sum_{k>i}
D^\phi_{ki}\,
\Delta_{ki}^{\phi}
-\sum_{j<i}
D^\phi_{ij}\,
\Delta_{ij}^{\phi},
\\[2mm]
\frac{dN_{B-L}}{dz}
&=
\sum_i
\epsilon_i D_i
\left(N_{N_i}-N_{N_i}^{\rm eq}\right)
-
\sum_i W_i N_{B-L}.
\end{align}
Here
\begin{equation}
\Delta_{ij}^{\phi}
\equiv
N_{N_i}
-
N_{N_i}^{\rm eq}
\frac{N_{N_j}}{N_{N_j}^{\rm eq}}
\frac{N_{\phi}}{N_{\phi}^{\rm eq}},
\qquad i>j,
\end{equation}
accounts for the departure from equilibrium of the scalar-mediated decay and
inverse-decay processes $N_i\leftrightarrow N_j\phi$. Throughout this
analysis we use the dimensionless time-evolution variable $z=M_2/T$. The
standard decay and washout terms take the usual form
\cite{Moffat_2018},
\begin{equation}
D_i(z)= K_i\,x_i\,z\,
\frac{\mathcal{K}_1(z_i)}{\mathcal{K}_2(z_i)},
\qquad
W_i(z)=\frac{1}{4}
K_i\sqrt{x_i}\,
z_i^3\mathcal{K}_1(z_i),
\end{equation}
where $\mathcal{K}_n$ denotes the modified Bessel function and
\begin{equation}
x_i=\frac{M_i^2}{M_2^2},
\qquad
z_i=\sqrt{x_i}\,z=\frac{M_i}{T},
\qquad
K_i=\frac{\Gamma_i^L}{H(T=M_i)}.
\end{equation}
The corresponding scalar-mediated decay term is defined analogously as
\begin{equation}
D^\phi_{ij}(z)
=
K^\phi_{ij}\,x_i\,z\,
\frac{\mathcal{K}_1(z_i)}{\mathcal{K}_2(z_i)},
\qquad
K^\phi_{ij}=\frac{\Gamma^\phi_{ij}}{H(T=M_i)}.
\end{equation}
The equilibrium comoving number densities for fermionic and bosonic
species are
\begin{equation}
N_f^{\rm eq}(z_i)
=
\frac{3}{8}z_i^2\mathcal{K}_2(z_i),
\qquad
N_b^{\rm eq}(z_i)
=
\frac{3}{16}z_i^2\mathcal{K}_2(z_i).
\end{equation}
The decay width for the process
$N_i\rightarrow N_j\phi$ is
\begin{equation}
\Gamma^\phi_{ij}
=
\frac{|y_{ij}|^2M_i}{16\pi}
\sqrt{\lambda(1,a_j,a_\phi)}
\left(
1-a_\phi+a_j+2\sqrt{a_j}
\right),
\end{equation}
where
\begin{equation}
a_j=\frac{M_j^2}{M_i^2},
\qquad
a_\phi=\frac{M_\phi^2}{M_i^2}.
\end{equation}
The scalar interaction also generates $2\leftrightarrow2$ scattering
processes involving the right-handed neutrinos. Their reaction densities
are higher order in the scalar couplings, scaling schematically as
$\gamma_S\propto |y_{ij}|^4$, whereas the scalar decay and inverse-decay
rates scale as $\Gamma^\phi_{ij}\propto |y_{ij}|^2$. Since our analysis is restricted to small scalar couplings, we work in the
decay--inverse-decay approximation and neglect these subleading scattering
processes.

\begin{figure}[h!]
\centering
\includegraphics[width=0.7\linewidth]{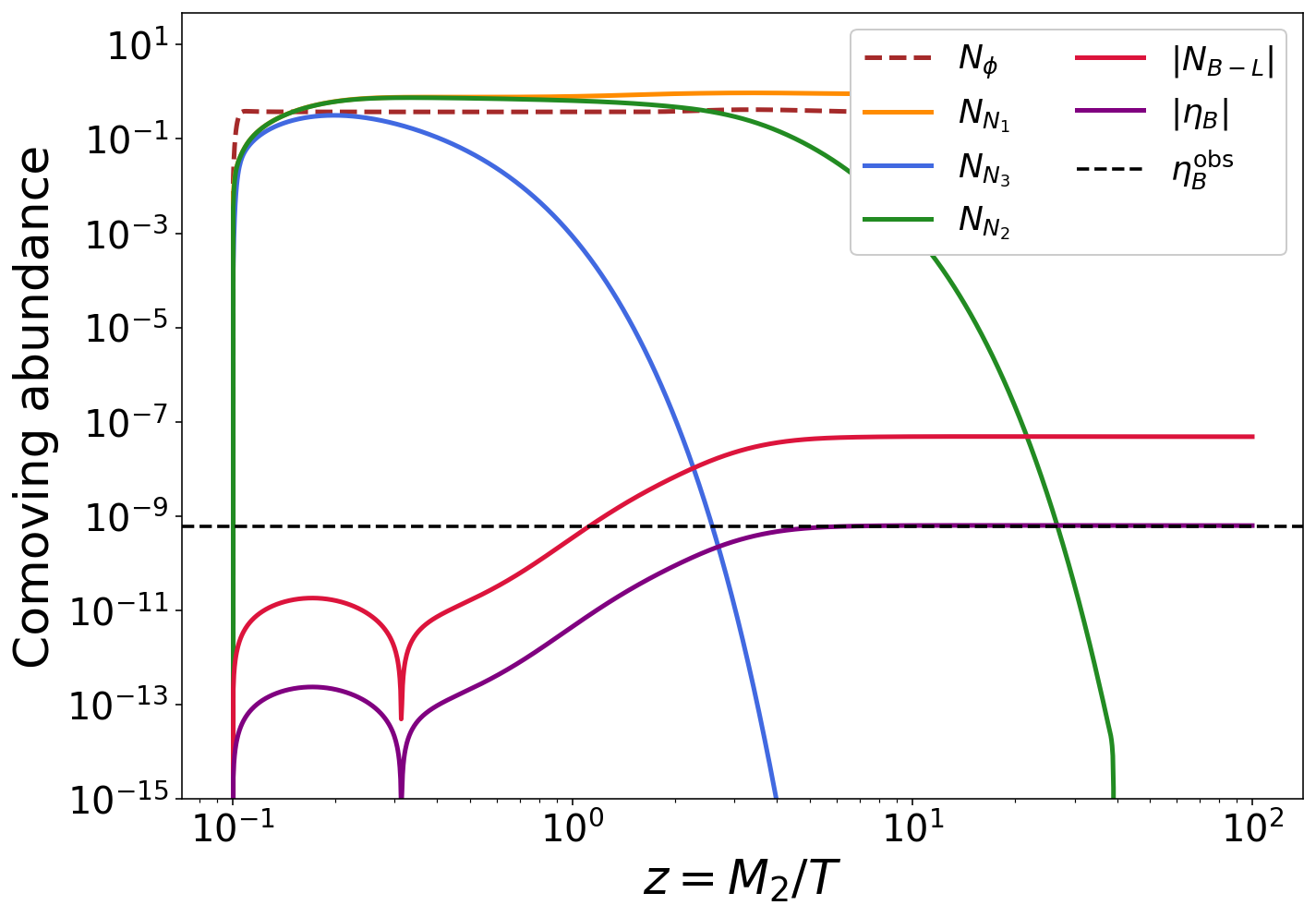}
\caption{\it Evolution of the comoving abundances for a representative
unflavoured benchmark with
$[M_1,M_2,M_3]=[10^{4},10^5,10^6]\,\mathrm{GeV}$ and
$M_\phi=2\times10^{3}\,\mathrm{GeV}$. We take
$\tilde m_1,\tilde m_2,\tilde m_3=
10^{-5},10^{-3},10^{-2}\,\mathrm{eV}$ and scalar couplings
$y_{21}=y_{13}=y_{32}=2.6\times10^{-3}$.
The CP asymmetry is taken to saturate the combined standard Type-I and
scalar-induced upper bounds. The dashed line denotes the observed
baryon-to-photon ratio.}
\label{fig:flavourless_bench}
\end{figure}

Finally, the generated $B-L$ asymmetry is converted into a baryon
asymmetry through electroweak sphaleron processes. With the abundance
normalisation adopted above, including the subsequent change in the
effective number of entropy degrees of freedom gives
\cite{Ulysses,Ulysses2}
\begin{equation}
\eta_B\simeq0.013\,N_{B-L}.
\end{equation}
In this work, we assume that the early Universe underwent a period of inflation, so that any pre-existing abundances of the right-handed neutrinos, the scalar, and the $B-L$ asymmetry are diluted away. We therefore adopt vanishing initial abundances,
\begin{equation}
N_\phi(z_{\rm in})
=
N_{N_i}(z_{\rm in})
=
N_{B-L}(z_{\rm in})
=
0.
\end{equation}
The initial temperature is taken sufficiently above the relevant
right-handed-neutrino masses such that the subsequent evolution is
insensitive to the precise choice of the starting temperature. We now present a representative benchmark illustrating the evolution of the
system in Fig.~\ref{fig:flavourless_bench}.

For this benchmark, we choose a
hierarchical RHN spectrum and a light scalar. Starting from vanishing initial
abundances, the RHNs are first produced from the thermal bath. The $N_3$
abundance reaches its maximum at earlier times before rapidly decaying, while
$N_2$ remains populated until $z=\mathcal{O}(10)$ and subsequently decays
away. The scalar-mediated decays simultaneously populate $N_1$ and $\phi$,
whose abundances closely track one another for this benchmark. Since the
coupling of $N_1$ to the Standard Model is suppressed, it does not
significantly wash out the asymmetry produced by the heavier states.

The $B-L$ asymmetry is generated predominantly during the departure of
$N_2$ from equilibrium. The dip visible in $|N_{B-L}|$ corresponds to a
change of sign of the evolving asymmetry before it approaches its final
value. After the washout processes become ineffective, the asymmetry freezes
out and gives a baryon-to-photon ratio in agreement with the observed value.
This benchmark therefore illustrates explicitly how the additional
scalar-induced CP asymmetry can produce successful leptogenesis for a
hierarchical RHN spectrum at a scale well below that of conventional
high-scale thermal leptogenesis. The required scalar couplings remain
perturbative and does not have to be unnaturally large. \\
We finally comment on the late-time fate of $N_1$ and $\phi$, whose
abundances remain non-negligible over the interval shown in
Fig.~\ref{fig:flavourless_bench}. Throughout the numerical analysis we take
$m_1\simeq0$, with sufficiently small but non-zero $Y_{\alpha1}$ such that
$N_1$ does not wash out the asymmetry generated by the heavier states but may
decay at later times. Likewise, a sufficiently small Higgs-portal interaction
may allow $\phi$ to decay into Standard Model states without affecting the
leptogenesis dynamics considered here. These late-time interactions are
independent of the mechanism responsible for generating the baryon asymmetry
and are not considered further in this work.\\
\subsection{Washout}
\begin{figure}[htbp]
    \centering

    \begin{minipage}[t]{0.31\textwidth}
        \centering
        \begin{tikzpicture}
            \begin{feynman}

                \vertex (v1) at (0,0);
                \vertex (v2) at (2.0,0);

                \vertex (L)   at (-1.4,0.7)  {\(L_\alpha\)};
                \vertex (H)   at (-1.4,-0.7) {\(H\)};
                \vertex (Nj)  at (3.4,0.7)   {\(N_j\)};
                \vertex (phi) at (3.4,-0.7)  {\(\phi\)};

                \diagram*{
                    (L) -- [fermion] (v1),
                    (H) -- [scalar] (v1),

                    (v1) -- [edge label=\(N_i\)] (v2),

                    (v2) -- (Nj),
                    (v2) -- [scalar] (phi),
                };

            \end{feynman}
        \end{tikzpicture}

        \vspace{2mm}

        \(\text{(a)}\quad
        L_\alpha H \leftrightarrow N_j\phi\)
    \end{minipage}
    \hfill
    \begin{minipage}[t]{0.31\textwidth}
        \centering
        \begin{tikzpicture}
            \begin{feynman}

                \vertex (v1) at (0,0.75);

                \vertex (v2) at (0,-0.75);

                \vertex (L)   at (-1.5,1.25) {\(L_\alpha\)};
                \vertex (phi) at (-1.5,-1.25) {\(\phi\)};

                \vertex (H)  at (1.5,1.25) {\(H^\dagger\)};
                \vertex (Nj) at (1.5,-1.25) {\(N_j\)};

                \diagram*{
                    (L) -- [fermion] (v1),
                    (v1) -- [scalar] (H),

                    (v1) -- [edge label'=\(N_i\)] (v2),

                    (phi) -- [scalar] (v2),
                    (v2) -- (Nj),
                };

            \end{feynman}
        \end{tikzpicture}

        \vspace{2mm}

        \(\text{(b)}\quad
        L_\alpha\phi \leftrightarrow H^\dagger N_j\)
    \end{minipage}
    \hfill
    \begin{minipage}[t]{0.31\textwidth}
        \centering
        \begin{tikzpicture}
            \begin{feynman}

                \vertex (v1) at (0,0.75);

                \vertex (v2) at (0,-0.75);

                \vertex (L)  at (-1.5,1.25)  {\(L_\alpha\)};
                \vertex (Nj) at (-1.5,-1.25) {\(N_j\)};

                \vertex (H)   at (1.5,1.25)  {\(H^\dagger\)};
                \vertex (phi) at (1.5,-1.25) {\(\phi\)};

                \diagram*{
                    (L) -- [fermion] (v1),
                    (v1) -- [scalar] (H),

                    (v1) -- [edge label'=\(N_i\)] (v2),

                    (Nj) -- (v2),
                    (v2) -- [scalar] (phi),
                };

            \end{feynman}
        \end{tikzpicture}

        \vspace{2mm}

        \(\text{(c)}\quad
        L_\alpha N_j \leftrightarrow H^\dagger\phi\)
    \end{minipage}

    \caption{\it
    New washout scatterings generated by one neutrino
    Yukawa interaction \(Y_{\alpha i}\) and one scalar interaction
    \(y_{ij}\).
    }
    \label{fig:scalar-assisted-washout}

\end{figure}
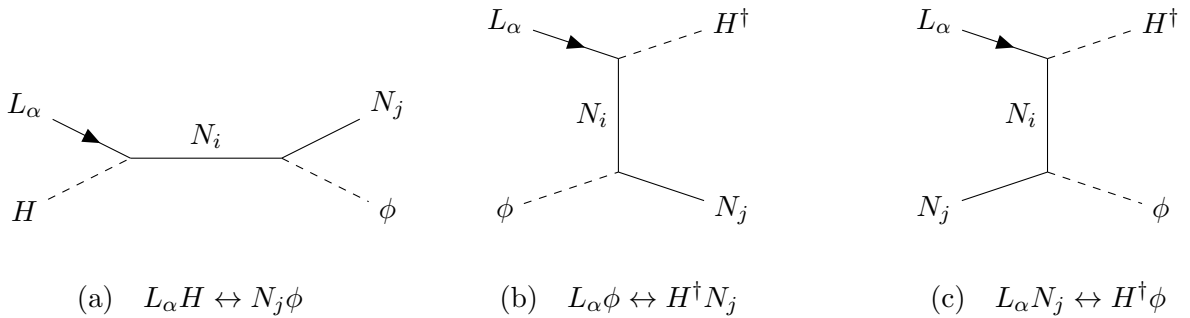

The additional scalar interaction generates new washout scatterings involving
one Standard Model Yukawa vertex and one scalar Yukawa vertex. The relevant
tree-level processes are shown in Fig.~\ref{fig:scalar-assisted-washout}.

We argue that the additional washout processes can be neglected in the parameter region considered in this work. To do this we consider separately the intermediate state being on and off-shell.

\begin{itemize}
    \item \textbf{On-shell $N_i$:} For the $s$-channel process $L_\alpha H \rightarrow N_i \rightarrow N_j\phi$, the intermediate right-handed neutrino may become on shell. In this case the process factorises into the inverse decay $L_\alpha H \rightarrow N_i$ followed by the scalar-mediated decay $N_i \rightarrow N_j\phi$. Both processes are already included separately in the Boltzmann equations, so including the corresponding real-intermediate-state contribution as an additional $2\leftrightarrow2$ scattering would constitute double counting and must therefore be subtracted. For the crossed processes $L_\alpha\phi \leftrightarrow H^\dagger N_j$ and $L_\alpha N_j \leftrightarrow H^\dagger\phi$, an on-shell intermediate $N_i$ is not kinematically accessible, so no real-intermediate-state subtraction is required.

    \item \textbf{Off-shell $N_i$:} After subtraction of the on-shell contribution, the $s$-channel process contains a genuinely new off-shell washout contribution, while the crossed diagrams are entirely off shell. Each of these processes contains one Standard Model Yukawa vertex and one scalar Yukawa vertex, so the amplitude scales as $\mathcal{M}_{\rm new}\propto Y_{\alpha i}y_{ij}$ and the corresponding reaction density scales schematically as $\gamma_{\rm new}\propto |Y_{\alpha i}|^2|y_{ij}|^2$. By comparison, the dominant standard inverse-decay washout scales as $W_i\propto (Y^\dagger Y)_{ii}$. The new washout contributions are therefore additionally suppressed by a factor of order $|y_{ij}|^2$. Since we consider $|y_{ij}|\ll1$ throughout this work, these off-shell contributions are sub-leading and can be neglected.
\end{itemize}

To summarise, the on-shell component of the $s$-channel process is already included in the decay--inverse-decay treatment, while the genuinely new off-shell $s$-channel and crossed contributions are suppressed by a factor of order $|y_{ij}|^2$ compared with the standard washout term. 
\section{Fully Flavoured Leptogenesis}
\label{sec:flavourful}

We now turn to the fully flavoured regime, where the charged-lepton Yukawa
interactions efficiently resolve the individual $e$, $\mu$, and $\tau$
flavours. This regime is particularly relevant for the low-scale scenarios
of interest in this work. We first formulate the corresponding flavoured
Boltzmann equations before presenting numerical results demonstrating
the impact of flavour on this mechanism.

\subsection{Flavoured Boltzmann Equations}

In the fully flavoured regime, the evolution equations for the scalar and
right-handed-neutrino abundances remain unchanged, while the total $B-L$
asymmetry is resolved into the three individual charged-lepton flavours.
The system is therefore given by
\begin{align}
\frac{dN_{\phi}}{dz}
&=
\sum_{i>j}
D^\phi_{ij}\,
\Delta_{ij}^{\phi},
\\[2mm]
\frac{dN_{N_i}}{dz}
&=
-D_i\left(N_{N_i}-N_{N_i}^{\rm eq}\right)
+\sum_{k>i}
D^\phi_{ki}\,
\Delta_{ki}^{\phi}
-\sum_{j<i}
D^\phi_{ij}\,
\Delta_{ij}^{\phi},
\\[2mm]
\frac{dN_{B-L}^{\alpha}}{dz}
&=
\sum_i
\epsilon^i_{\alpha}D_i
\left(N_{N_i}-N_{N_i}^{\rm eq}\right)
-
\sum_i
W_iP^i_{\alpha\alpha}
N_{B-L}^{\alpha},
\qquad
\alpha=e,\mu,\tau .
\end{align}
All quantities entering the equations for $N_\phi$ and $N_{N_i}$ are
defined as in the previous section. The only modification is the resolution
of the CP-violating source and washout into flavour space. The flavoured
CP asymmetries $\epsilon^i_\alpha$ include both the standard Type-I and
scalar-induced contributions derived in Section~\ref{sec:model}. The
tree-level flavour projector is defined by
\begin{equation}
P^i_{\alpha\alpha} =\frac{\widetilde m^i_{\alpha}}{\widetilde m_i}=
\frac{|Y_{\alpha i}|^2}{(Y^\dagger Y)_{ii}},
\end{equation}
where
\begin{equation}
\widetilde m^i_{\alpha}
=
\frac{v^2|Y_{\alpha i}|^2}{M_i},
\qquad
\widetilde m_i
=
\sum_\alpha \widetilde m^i_\alpha
=
\frac{v^2(Y^\dagger Y)_{ii}}{M_i}.
\end{equation}
The projectors determine the fraction of the inverse-decay washout
associated with each charged-lepton flavour. The total lepton and baryon asymmetry is recovered
by summing over flavours,
\begin{equation}
N_{B-L}
=
\sum_{\alpha=e,\mu,\tau}N_{B-L}^{\alpha},\qquad \eta_B
\simeq
0.013
\sum_{\alpha=e,\mu,\tau}N_{B-L}^{\alpha}.
\end{equation}
As in the unflavoured analysis, we take vanishing initial abundances for the
right-handed neutrinos, the scalar, and the flavoured $B-L$ asymmetries,
\begin{equation}
N_{N_i}(z_{\rm in})
=
N_{\phi}(z_{\rm in})
=
N_{B-L}^{\alpha}(z_{\rm in})
=
0.
\end{equation}
For the numerical analyses below, we restrict the complex orthogonal matrix
to a rotation in the $2$--$3$ plane, parametrised by the complex angle
$\omega_{23}$,
\begin{equation}
R=
\begin{pmatrix}
1 & 0 & 0\\
0 & \cos\omega_{23} & -\sin\omega_{23}\\
0 & \sin\omega_{23} & \cos\omega_{23}
\end{pmatrix}
\end{equation}
where $\omega_{23}$ is in general complex. This choice leaves $N_1$ aligned with the lightest-neutrino direction and therefore
suppresses its Standard Model Yukawa couplings, placing $N_1$ in the
weak-washout regime.

\subsection{Results}
\begin{figure}[h!]
\centering
\includegraphics[width=1\linewidth]{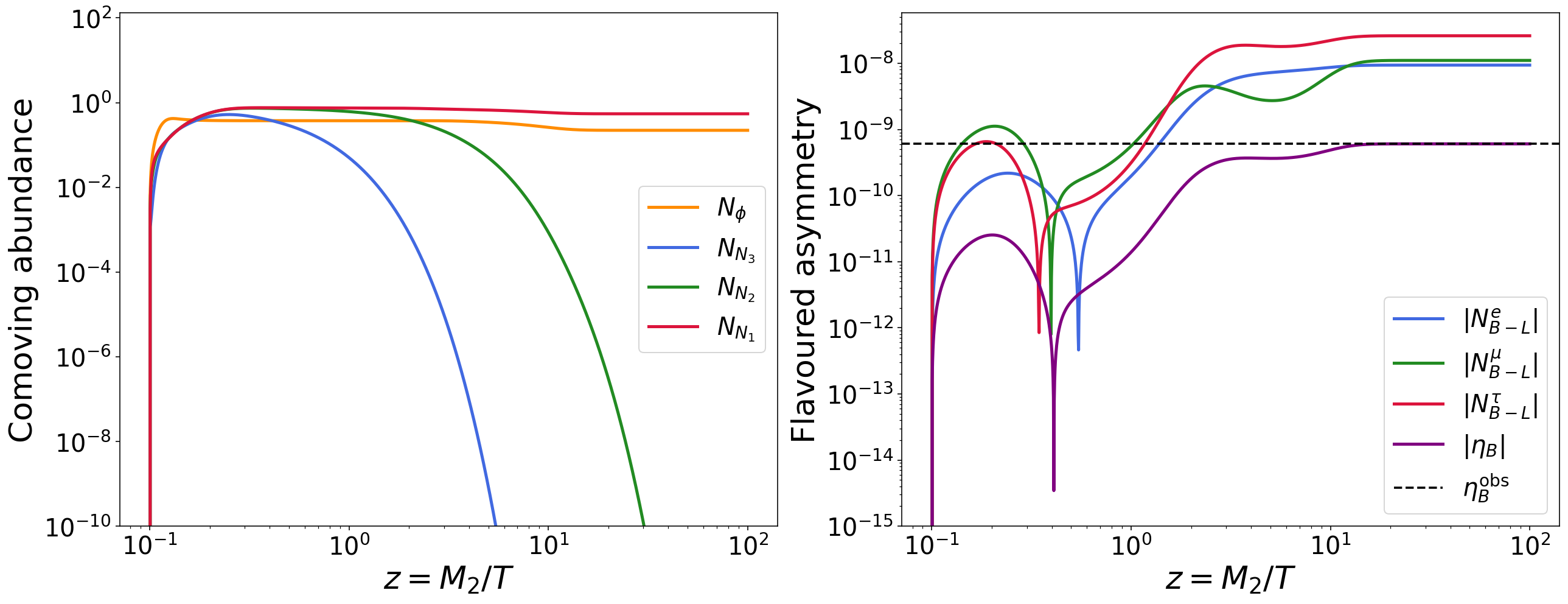}
\caption{\it
Evolution of the comoving abundances and flavoured asymmetries for a
representative fully flavoured benchmark. The left panel shows the scalar
and right-handed-neutrino abundances, $N_\phi$, $N_{N_1}$, $N_{N_2}$, and
$N_{N_3}$, while the right panel shows the individual asymmetries
$|N^e_{B-L}|$, $|N^\mu_{B-L}|$, and $|N^\tau_{B-L}|$, together with the
resulting baryon-to-photon ratio $|\eta_B|$. We take
$M_1=2\times10^{6}\,{\rm GeV}$, $M_2=10^{7}\,{\rm GeV}$,
$M_3=5\times10^{7}\,{\rm GeV}$, and $M_\phi=2\times10^{5}\,{\rm GeV}$,
with scalar couplings $y_{21}=y_{13}=y_{32}=6.0\times10^{-3}$.
The light-neutrino sector assumes normal ordering with $m_1=0$ and the
loop-corrected Casas--Ibarra angle $\omega_{23}=0.62+0.20i$.
The dashed horizontal line denotes the observed baryon-to-photon ratio.}
\label{fig:flavoured_bench}
\end{figure}

We first consider a representative benchmark in the fully flavoured regime.
The purpose of this benchmark is to demonstrate that the mechanism remains effective for a hierarchical right-handed-neutrino spectrum while keeping the scalar couplings perturbative. We construct the neutrino Yukawa matrix explicitly using the loop-corrected Casas--Ibarra
parametrisation, so that the flavour projectors, washout parameters, and CP
asymmetries entering the Boltzmann equations are derived consistently from
the same underlying neutrino sector. The resulting evolution is shown in Fig.~\ref{fig:flavoured_bench}.

The three flavour asymmetries evolve differently as a consequence of the
flavour dependence of both the CP-violating source and the inverse-decay
washout. Once the relevant inverse-decay processes become inefficient, the individual
flavour asymmetries freeze out and their sum determines the final baryon
asymmetry. For the benchmark shown in Fig.~\ref{fig:flavoured_bench}, the
observed asymmetry is reproduced for a hierarchical heavy-neutrino spectrum
at a scale far below that usually associated with conventional hierarchical
Type-I leptogenesis.

At intermediate scales, however, reproducing the baryon asymmetry is not by itself
sufficient to demonstrate that the mechanism avoids the tuning associated
with low-scale leptogenesis in the minimal Type-I seesaw. Increasing the
ordinary neutrino Yukawa couplings can enhance the CP asymmetry, but can
simultaneously increase the radiative contribution to the light-neutrino
mass matrix, leading to cancellations between the tree-level and one-loop
terms. We therefore quantify this cancellation explicitly for our benchmark. Following Ref.~\cite{Moffat_2018}, we write the physical light-neutrino mass
matrix as $m_\nu=m_\nu^{\rm tree}+m_\nu^{\rm 1-loop}$ and quantify the
relative size of the radiative contribution by
\begin{equation}
{\rm F.T.}
\equiv
\frac{\displaystyle\sum_{i=1}^{3}
{\rm SVD}\!\left[m_\nu^{\rm 1-loop}\right]_i}
{\displaystyle\sum_{i=1}^{3}
{\rm SVD}\!\left[m_\nu\right]_i},
\end{equation}
where ${\rm SVD}[m]_i$ denotes the $i$th singular value of the corresponding
mass matrix. In the tree-dominated limit the radiative correction is small
and ${\rm F.T.}\rightarrow0$, while increasing values indicate a progressively
more important one-loop contribution and, eventually, a cancellation between
the tree-level and radiative terms. For the parameter point shown in Fig.~\ref{fig:flavoured_bench}, we find
${\rm F.T.}\simeq0.12$. The light-neutrino mass matrix therefore remains
tree dominated. The observed baryon asymmetry is not obtained by enhancing
the ordinary neutrino Yukawa couplings and tuning their tree-level
contribution against radiative corrections. Instead, the additional CP
violation required for leptogenesis is supplied by the scalar sector.

This illustrates the central advantage of the singlet-scalar extension:
the additional scalar interaction partially decouples the size of the
CP-violating source from the Yukawa couplings responsible for the
light-neutrino masses. Successful intermediate-scale leptogenesis can therefore be obtained without relying on a large tree--loop cancellation in the active
neutrino sector.

At sufficiently low leptogenesis scales, electroweak sphaleron freeze-out becomes important. In the preceding benchmark the asymmetry is generated well above the electroweak scale, so the final baryon asymmetry may be obtained
directly from $\eta_B\simeq0.013\,N_{B-L}$. When $M_2$ approaches the
electroweak scale, however, a significant fraction of the lepton asymmetry
may be generated after sphaleron processes have become ineffective and can
therefore no longer contribute to the baryon asymmetry. We take the sphaleron freeze-out temperature to be
$T_{\rm sph}=131.7\,{\rm GeV}$ and approximate sphaleron decoupling as
instantaneous. The right-handed-neutrino, scalar, and flavoured $B-L$
abundances are evolved normally through and beyond this point, while the
baryon asymmetry is evaluated at $z=z_{\rm sph}$ and subsequently held fixed,
\begin{equation}
\eta_B(z\geq z_{\rm sph})
=
0.013\sum_{\alpha=e,\mu,\tau}N^\alpha_{B-L}(z_{\rm sph}).
\end{equation}
This allows us to determine whether this mechanism remains
viable when the characteristic heavy-neutrino scale is lowered to the
electroweak scale.

\begin{figure}[h!]
\centering
\includegraphics[width=1\linewidth]{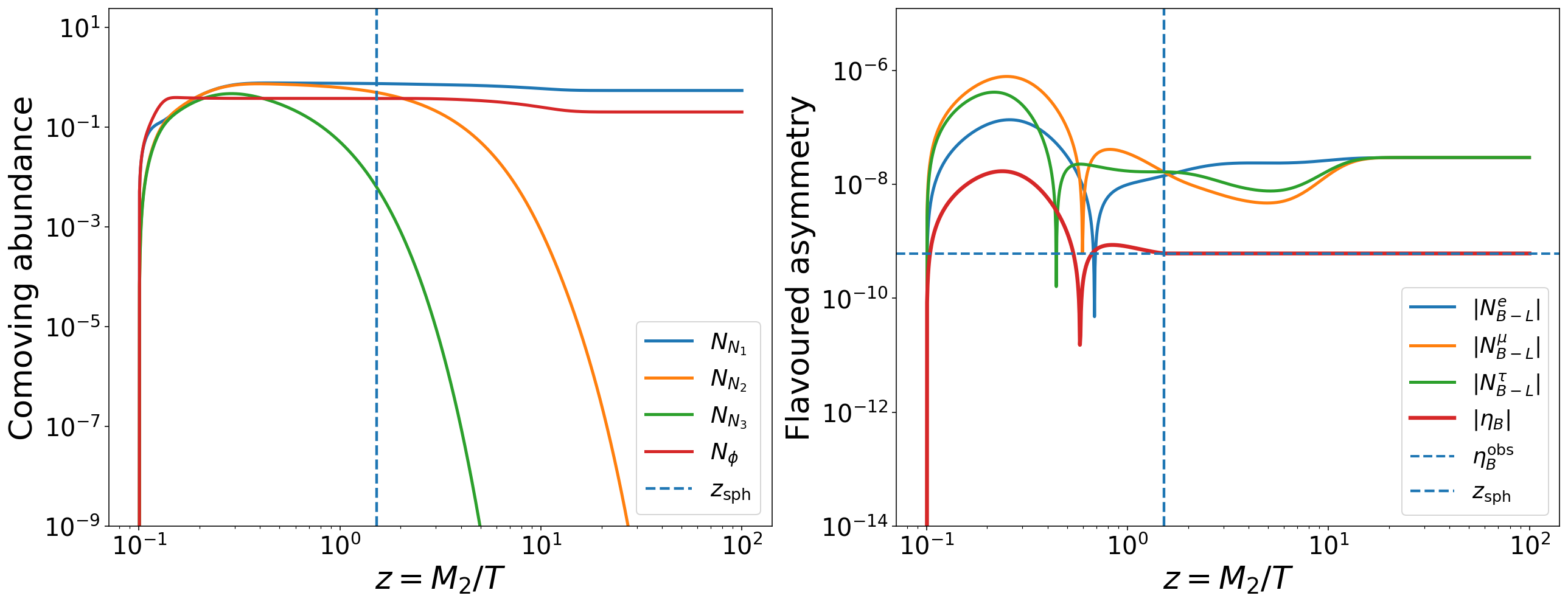}
\caption{\it
Evolution of the comoving abundances and flavoured asymmetries for an
electroweak-scale leptogenesis benchmark with
$(M_\phi,M_1,M_2,M_3)=(4,40,200,1000)\,{\rm GeV}$.
We take $y_{21}=3.25\times10^{-3}$,
$y_{13}=3\times10^{-2}$, $y_{32}=5\times10^{-6}$, and
$\omega_{23}=1.20-0.40i$.
The left panel shows the abundances $N_{N_1}$, $N_{N_2}$, $N_{N_3}$, and
$N_\phi$. The right panel shows the individual flavoured asymmetries
$|N^e_{B-L}|$, $|N^\mu_{B-L}|$, and $|N^\tau_{B-L}|$, together with the
baryon-to-photon ratio $|\eta_B|$. The vertical dashed line denotes
sphaleron freeze-out, $z_{\rm sph}=M_2/T_{\rm sph}\simeq1.52$, for
$T_{\rm sph}=131.7\,{\rm GeV}$. The $B-L$ asymmetries continue to evolve
after this point, while $\eta_B$ is fixed at its value at $z_{\rm sph}$,
since subsequent lepton-number production can no longer be converted into
baryon number. The horizontal dashed line denotes the observed
baryon-to-photon ratio.}
\label{fig:low_scale_benchmark}
\end{figure}

The electroweak-scale benchmark avoids the ingredients commonly required
to lower the scale of leptogenesis in the minimal Type-I seesaw. The
right-handed-neutrino spectrum remains hierarchical, with $M_2=5M_1$ and
$M_3=5M_2$, and therefore does not rely on resonant enhancement from
quasi-degenerate right-handed neutrinos. The required scalar couplings remain
perturbative, while the neutrino-mass fine-tuning measure is
${\rm F.T.}=0.01$, showing that the light-neutrino mass
matrix remains strongly tree dominated. The additional CP violation required
for leptogenesis is instead supplied by the scalar interaction.

This benchmark therefore provides an explicit example of hierarchical
thermal leptogenesis at the electroweak scale without requiring either
quasi-degenerate right-handed-neutrino masses or a large cancellation between
the tree-level and one-loop contributions to the active-neutrino masses.
Moreover, because the right-handed neutrinos remain close to the electroweak
scale, the large radiative corrections to the Higgs mass associated with
conventional high-scale leptogenesis are avoided.

\section{Density-Matrix Formalism}
\label{sec:DME}

The unflavoured and fully flavoured Boltzmann equations considered above
correspond to two limiting descriptions of the lepton-flavour dynamics. In
the intermediate flavour regime, however, the charged-lepton Yukawa
interactions are only partially effective and the lepton state produced in
RHN decays retains non-negligible flavour coherence. In this regime the
asymmetry must be described by a density matrix in flavour space
\cite{Moffat_2018,Ulysses,Ulysses2}. We therefore formulate the corresponding
density-matrix equations for this model.

\subsection{Density-Matrix Equations}

The equations governing the RHN and scalar abundances remain unchanged,
while the single quantity $N_{B-L}$ is promoted to the Hermitian matrix
$N^{B-L}_{\alpha\beta}$ and the evolution is described by
\begin{align}
\frac{dN^{B-L}_{\alpha\beta}}{dz}
={}&
\sum_i
\epsilon_{\alpha\beta}^{(i)}
D_i\left(N_{N_i}-N_{N_i}^{\rm eq}\right)
-
\frac{1}{2}
\sum_i W_i
\left\{
P^{0(i)},N^{B-L}
\right\}_{\alpha\beta}
\nonumber\\
&-
\frac{\Lambda_\tau}{Hz}
\left[
\Pi_\tau,
\left[
\Pi_\tau,N^{B-L}
\right]
\right]_{\alpha\beta}
-
\frac{\Lambda_\mu}{Hz}
\left[
\Pi_\mu,
\left[
\Pi_\mu,N^{B-L}
\right]
\right]_{\alpha\beta},
\label{eq:density_matrix}
\end{align}
where the tree-level flavour projector associated with $N_i$ decay is
\begin{equation}
P_{\alpha\beta}^{0(i)}
=
\frac{Y_{\alpha i}Y_{\beta i}^{*}}
{(Y^\dagger Y)_{ii}}.
\end{equation}
The charged-lepton projectors appearing in
Eq.~\eqref{eq:density_matrix} are
\begin{equation}
\Pi_\tau=
\begin{pmatrix}
0&0&0\\
0&0&0\\
0&0&1
\end{pmatrix},
\qquad
\Pi_\mu=
\begin{pmatrix}
0&0&0\\
0&1&0\\
0&0&0
\end{pmatrix},
\end{equation}
and the corresponding interaction rates may be approximated by \cite{Moffat_2018}
\begin{equation}
\Lambda_\alpha\simeq
8\times10^{-3}\,h_\alpha^2 T,
\qquad
h_\alpha=\frac{m_\alpha}{v}.
\end{equation}
The double commutators in Eq.~\eqref{eq:density_matrix} damp the
off-diagonal components of $N^{B-L}$ while leaving its diagonal components
unchanged, thereby describing the progressive decoherence of the lepton
state as the charged-lepton Yukawa interactions enter equilibrium.

The CP asymmetry appearing in the density-matrix equation is itself a
Hermitian matrix,
\begin{equation}
\epsilon_{\alpha\beta}^{(i)}
=
\epsilon_{\alpha\beta}^{(i),{\rm Type-I}}
+
\epsilon_{\alpha\beta}^{(i),\phi}.
\label{eq:epsilon_matrix_total}
\end{equation}
Its diagonal components reproduce the flavoured CP asymmetries,
$\epsilon_{\alpha\alpha}^{(i)}=\epsilon_\alpha^i$, while its trace gives the
usual unflavoured asymmetry,
${\rm Tr}[\epsilon^{(i)}]=\sum_\alpha\epsilon_\alpha^i=\epsilon_i$.
The off-diagonal components contain the additional information associated
with coherent superpositions of charged-lepton flavours. For the standard Type-I contribution, the CP-asymmetry matrix can be written
as
\begin{align}
\epsilon_{\alpha\beta}^{(i),{\rm Type-I}}
={}&
\frac{3}{32\pi (Y^\dagger Y)_{ii}}
\sum_{j\neq i}
\Bigg\{
i
\left[
Y_{\alpha i}Y_{\beta j}^{*}(Y^\dagger Y)_{ji}
-
Y_{\beta i}^{*}Y_{\alpha j}(Y^\dagger Y)_{ij}
\right]
\frac{\xi(x_{ji})}{\sqrt{x_{ji}}}
\nonumber\\
&\hspace{1.0cm}
+
\frac{2i}{3(x_{ji}-1)}
\left[
Y_{\alpha i}Y_{\beta j}^{*}(Y^\dagger Y)_{ij}
-
Y_{\beta i}^{*}Y_{\alpha j}(Y^\dagger Y)_{ji}
\right]
\Bigg\},
\label{eq:epsilon_matrix_typeI}
\end{align}
where $x_{ji}=M_j^2/M_i^2$ and
\begin{equation}
\xi(x)
=
\frac{2}{3}x
\left[
(1+x)\ln\left(\frac{1+x}{x}\right)
-
\frac{2-x}{1-x}
\right].
\end{equation}

The scalar interaction introduces an additional contribution to the
CP-asymmetry matrix. Generalising the flavoured result derived in
Section~\ref{sec:model}, this contribution takes the form
\begin{align}
\epsilon_{\alpha\beta}^{(i),\phi}
={}&
\frac{1}{4\pi(Y^\dagger Y)_{ii}}
\sum_{\substack{j<i\\k\neq i}}
\frac{M_i+M_k}{M_i^2-M_k^2}
f(M_i^2,M_j^2,M_\phi^2)
\nonumber\\
&\times
\frac{1}{2i}
\left[
Y_{\alpha k}Y_{\beta i}^*\,y_{ij}y_{jk}
-
Y_{\alpha i}Y_{\beta k}^*\,y_{ij}^*y_{jk}^*
\right].
\label{eq:scalar_cp_density}
\end{align}

For the $N_2$-dominated scenario considered throughout this work, the
relevant scalar contribution arises from the intermediate $N_1$ state and
the virtual $N_3$ state. Equation~\eqref{eq:scalar_cp_density} therefore
reduces to
\begin{align}
\epsilon_{\alpha\beta}^{(2),\phi}
={}&
\frac{1}{4\pi(Y^\dagger Y)_{22}}
\frac{M_2+M_3}{M_2^2-M_3^2}
f(M_2^2,M_1^2,M_\phi^2)
\nonumber\\
&\times
\frac{1}{2i}
\left[
Y_{\alpha3}Y_{\beta2}^*\,y_{21}y_{13}
-
Y_{\alpha2}Y_{\beta3}^*\,y_{21}^*y_{13}^*
\right].
\end{align}
For real scalar couplings and in the hierarchical limit
$M_3\gg M_2\gg M_1,M_\phi$, this simplifies to
\begin{equation}
\epsilon_{\alpha\beta}^{(2),\phi}
\simeq
\frac{M_2}{16\pi(Y^\dagger Y)_{22}}
\frac{y_{21}y_{13}}{M_2-M_3}
\frac{1}{2i}
\left(
Y_{\alpha3}Y_{\beta2}^*
-
Y_{\alpha2}Y_{\beta3}^*
\right).
\end{equation}
Setting $\alpha=\beta$ reproduces the corresponding flavoured scalar
asymmetry, while taking the trace reproduces the unflavoured quantity
$\epsilon_2^\phi$.

\subsection{Benchmark}

\begin{figure}[h!]
\centering
\includegraphics[width=1\linewidth]{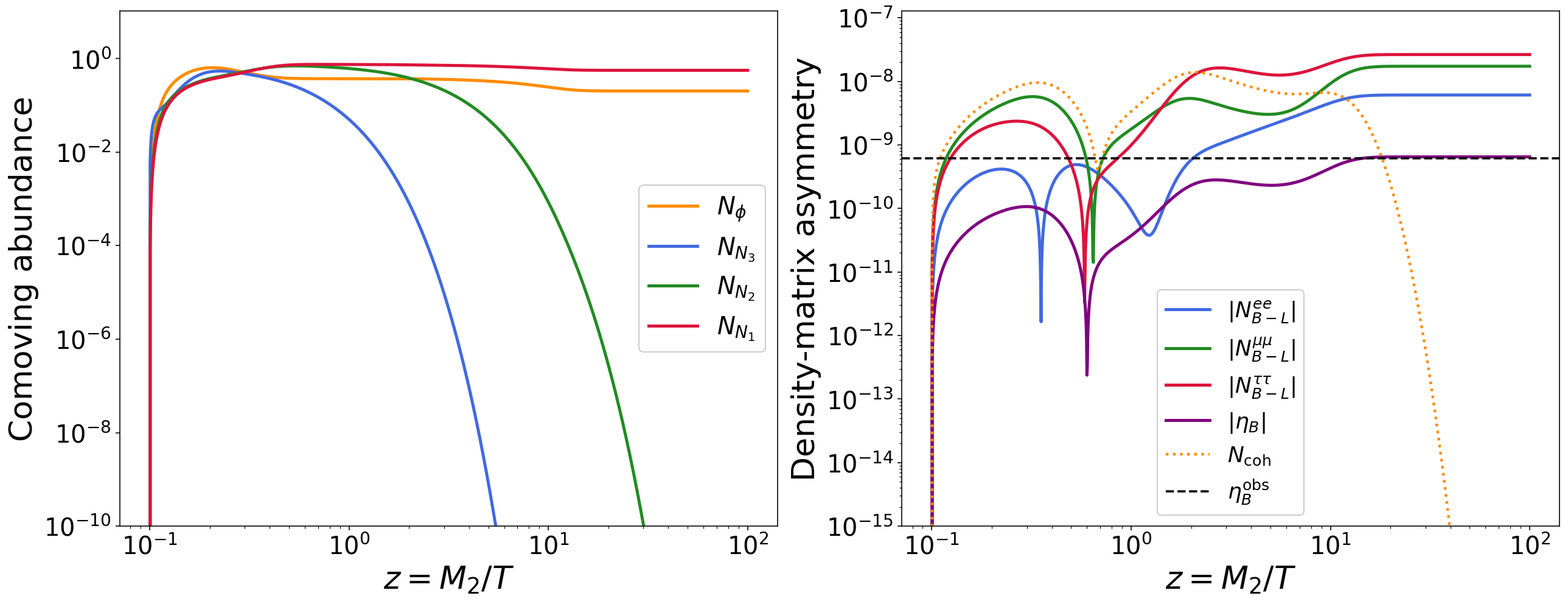}
\caption{\it
Evolution of the comoving abundances and density-matrix asymmetries for a
representative benchmark in the intermediate flavour regime. The left panel
shows the abundances $N_\phi$, $N_{N_1}$, $N_{N_2}$, and $N_{N_3}$, while the
right panel shows the diagonal components $|N^{ee}_{B-L}|$,
$|N^{\mu\mu}_{B-L}|$, and $|N^{\tau\tau}_{B-L}|$, together with the
baryon-to-photon ratio $|\eta_B|$. The dotted orange curve denotes the
magnitude of the off-diagonal flavour coherence,
$N_{\rm coh}=\sqrt{2\left(|N^{e\mu}_{B-L}|^2+|N^{e\tau}_{B-L}|^2+
|N^{\mu\tau}_{B-L}|^2\right)}$. We take
$(M_1,M_2,M_3,M_\phi)=(6\times10^8,3\times10^9,1.5\times10^{10},
6\times10^7)\,{\rm GeV}$, with
$y_{21}=y_{13}=y_{32}=10^{-2}$, normal ordering with $m_1=0$, and
$\omega_{23}=0.806+0.20i$. The dashed horizontal line denotes the observed
baryon-to-photon ratio.
}
\label{fig:DME_bench}
\end{figure}

We consider a representative benchmark in the intermediate flavour regime shown in Fig.~\ref{fig:DME_bench}.

Figure~\ref{fig:DME_bench} illustrates the evolution of the system when
flavour coherence is retained explicitly. The RHN and scalar abundances
follow the same qualitative behaviour as in the previous benchmarks, with
$N_3$ decaying first and the dominant $N_2$ dynamics occurring around
$z=\mathcal{O}(1)$. The scalar-mediated interactions populate both $N_1$ and
$\phi$, while the suppressed Standard Model coupling of $N_1$ prevents it
from generating significant leptonic washout.

The right panel demonstrates the additional information contained in the
density-matrix description. During the production of the asymmetry, the
off-diagonal components become sizeable, indicating that the lepton state
cannot yet be described as an incoherent statistical mixture of the three
charged-lepton flavours. As the Universe cools, charged-lepton Yukawa
interactions progressively destroy this coherence, and $N_{\rm coh}$ is
driven towards zero. The evolution therefore interpolates continuously
between the coherent and fully flavoured limits.

The diagonal components evolve differently as a consequence of their
distinct CP-violating source terms and washout rates. As in the previous
benchmarks, the sharp minima in the absolute asymmetries correspond to
changes of sign during their evolution. Once the washout and decoherence processes become
ineffective, the diagonal asymmetries freeze out and their trace determines
the final $B-L$ asymmetry. 

This benchmark demonstrates that the scalar-extended model remains viable
in the intermediate flavour regime when flavour coherence and decoherence are
treated consistently.

\section{Discussion \& Conclusion}
\label{sec:conclusion}

Minimal Type-I leptogenesis faces a tension between successful baryogenesis
and naturalness. For a hierarchical right-handed-neutrino spectrum,
conventional thermal leptogenesis typically requires a high mass scale, which
induces large radiative corrections to the Higgs mass parameter. Lowering the
leptogenesis scale within the minimal framework generally requires additional
structure, either through quasi-degenerate right-handed-neutrino masses for
resonant enhancement or through cancellations between the tree-level and
one-loop contributions to the active-neutrino mass matrix.

In this work, we have shown that extending the Type-I seesaw by a real singlet
scalar provides an alternative route to intermediate and low-scale leptogenesis. Off-diagonal scalar couplings generate an additional source of CP violation in the decays
of the heavier right-handed neutrinos without a corresponding enhancement of
the standard leptonic washout. We have derived an analytic upper bound on this
scalar-induced CP asymmetry, demonstrating explicitly its dependence on the
new scalar couplings and the right-handed-neutrino mass hierarchy.

Since the parameter region relevant for this mechanism generally lies outside
the single-flavour limit, we have incorporated flavour effects
using fully flavoured Boltzmann equations and, in the intermediate regime, the
density-matrix formalism. Explicit benchmark solutions demonstrate that the
observed baryon asymmetry can be generated with a hierarchical
right-handed-neutrino spectrum down to the electroweak scale.

The central result is therefore that scalar-extended leptogenesis can realise
low-scale hierarchical thermal leptogenesis without the tuning ingredients typically
required in the minimal Type-I seesaw.

\section*{Acknowledgements}
I wish to thank Pasquale Di Bari for many discussions on Leptogenesis and Graham White for discussions and comments on the manuscript. I acknowledge the STFC Consolidated Grant ST/X000583/1 and thank the University of Southampton School of Physics and Astronomy for the support of a Mayflower PhD scholarship.

\appendix
\section{CP-Asymmetry Calculation in Detail}
\label{sec:appendix}

\begin{figure}[h!]
\centering
\begin{tikzpicture}
  \begin{feynman}
    \vertex (a) at (-6,0);
    \vertex (b) at (-4,0);
    \vertex (c) at (-2,0);
    \vertex (d) at (0,0);
    \vertex (e) at (2,1);
    \vertex (f) at (2,-1);

    \diagram* {
      (a) -- [plain, edge label=\(N_i\)] (b),

      (b) -- [half left, scalar, looseness=1.3,
              edge label=\(\phi\)] (c),
      (b) -- [half right, plain, looseness=1.3,
              edge label=\(N_j\)] (c),

      (c) -- [plain, edge label=\(N_k\)] (d),

      (d) -- [fermion, edge label=\(\ell_\alpha\)] (e),
      (d) -- [scalar, edge label'=\(H\)] (f),
    };
  \end{feynman}
\end{tikzpicture}
\caption{\itshape
Scalar-induced one-loop contribution to the decay
$N_i\rightarrow \ell_\alpha H$ arising from the off-diagonal
couplings $y_{ij}$.}
\label{fig:appendix_cp_diagram}
\end{figure}
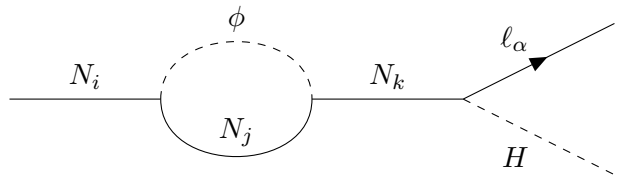

In this appendix we derive the scalar-induced contribution to the
CP asymmetry quoted in Section~\ref{sec:model}. The relevant one-loop diagram is shown in Fig.~\ref{fig:appendix_cp_diagram}.

\subsection{Flavoured and Unflavoured CP Asymmetries}

Following Ref.~\cite{Giudice_2004}, we write the tree-level and one-loop
amplitudes for a given charged-lepton flavour $\alpha$ as
\begin{equation}
\mathcal{M}_\alpha
=
\lambda_{0,\alpha} I_0
+
\lambda_{1,\alpha} I_1 .
\end{equation}
The CP-violating contribution from their interference is
\begin{equation}
\epsilon_{i\alpha}^{\phi}
=
-2\,
\frac{
\operatorname{Im}
\left(
\lambda_{0,\alpha}^{*}\lambda_{1,\alpha}
\right)}
{(Y^\dagger Y)_{ii}}
\frac{
\operatorname{Im}\left(I_0^{*}I_1\right)}
{|I_0|^2},
\label{eq:appendix_interference}
\end{equation}
where the denominator is normalised to the total tree-level decay width of
$N_i$. For fixed internal states $N_j$ and $N_k$, the coupling and kinematic
structures are
\begin{align}
\lambda_{0,\alpha}
&=
Y_{\alpha i},
\\
\lambda_{1,\alpha}
&=
-
y_{ij}y_{jk}Y_{\alpha k}
\frac{M_i+M_k}{M_i^2-M_k^2},
\\
I_0
&=
\bar{u}(q)P_Lu(p),
\\
I_1
&=
\bar{u}(q)P_L
\left[
\int\frac{d^4k}{(2\pi)^4}
\frac{\slashed{k}+M_j}
{\left(k^2-M_j^2\right)
\left[(p-k)^2-M_\phi^2\right]}
\right]
u(p).
\end{align}
Here the overall real sign in $\lambda_{1,\alpha}$ follows from the
Feynman-rule convention adopted for the scalar loop and ensures consistency
with the CP-asymmetry convention used in the main text. The factor
$(M_i+M_k)/(M_i^2-M_k^2)$ follows from the virtual $N_k$ propagator,
since
\begin{equation}
\frac{\slashed{p}+M_k}{p^2-M_k^2}u(p)
=
\frac{M_i+M_k}{M_i^2-M_k^2}u(p),
\end{equation}
where $\slashed{p}\,u(p)=M_i u(p)$ has been used.

The absorptive part of the loop is obtained using the Cutkosky cutting
rules. Cutting the internal $N_j$ and $\phi$ propagators amounts to the
replacement
\begin{equation}
\frac{1}{k^2-m^2+i\varepsilon}
\longrightarrow
-2\pi i\,
\delta(k^2-m^2)\theta(k^0).
\end{equation}
Applying this to both internal lines gives
\begin{align}
2\,\operatorname{Im}[I_1]
={}&
\bar{u}(q)P_L
\Bigg[
\int\frac{d^4k}{(2\pi)^4}
(\slashed{k}+M_j)
(2\pi)\delta(k^2-M_j^2)\theta(k^0)
\nonumber\\
&\hspace{2.0cm}\times
(2\pi)\delta((p-k)^2-M_\phi^2)
\theta(p^0-k^0)
\Bigg]
u(p).
\end{align}
We evaluate the integral in the rest frame of the decaying RHN,
\begin{equation}
p^\mu=(M_i,\mathbf{0}),
\qquad
k^\mu=(k^0,\mathbf{k}).
\end{equation}
Defining
\begin{equation}
E_j=\sqrt{|\mathbf{k}|^2+M_j^2},
\qquad
E_\phi=\sqrt{|\mathbf{k}|^2+M_\phi^2},
\end{equation}
the positive-energy delta functions impose $k^0=E_j$ and $M_i-E_j=E_\phi$. A non-zero absorptive contribution therefore requires
\begin{equation}
M_i>M_j+M_\phi.
\label{eq:appendix_threshold}
\end{equation}
The on-shell three-momentum and energies are
\begin{align}
|\mathbf{k}|_\star
&=
\frac{
\sqrt{\lambda(M_i^2,M_j^2,M_\phi^2)}
}{2M_i},
\\
E_j
&=
\frac{M_i^2+M_j^2-M_\phi^2}{2M_i},
\\
E_\phi
&=
\frac{M_i^2-M_j^2+M_\phi^2}{2M_i}.
\end{align}
Performing the $k^0$ integral gives
\begin{equation}
2\,\operatorname{Im}[I_1]
=
\bar{u}(q)P_L
\left[
\int\frac{d^3\mathbf{k}}{(2\pi)^2}
\frac{\slashed{k}+M_j}{4E_jE_\phi}
\delta(M_i-E_j-E_\phi)
\right]
u(p).
\end{equation}
To perform the radial integral, define
\begin{equation}
g(|\mathbf{k}|)
=
M_i-E_j-E_\phi .
\end{equation}
At $|\mathbf{k}|=|\mathbf{k}|_\star$,
\begin{equation}
\left|
\frac{dg}{d|\mathbf{k}|}
\right|_\star
=
|\mathbf{k}|_\star
\left(
\frac{1}{E_j}+\frac{1}{E_\phi}
\right)
=
\frac{|\mathbf{k}|_\star M_i}{E_jE_\phi},
\end{equation}
where $E_j+E_\phi=M_i$ has been used. Hence
\begin{equation}
\delta(M_i-E_j-E_\phi)
=
\frac{E_jE_\phi}
{|\mathbf{k}|_\star M_i}
\,
\delta\!\left(
|\mathbf{k}|-|\mathbf{k}|_\star
\right).
\end{equation}
Using $d^3\mathbf{k}=|\mathbf{k}|^2d|\mathbf{k}|\,d\Omega$, the radial
integration then gives
\begin{equation}
2\,\operatorname{Im}[I_1]
=
\frac{|\mathbf{k}|_\star}{4\pi M_i}
\left(E_j+M_j\right)
\bar{u}(q)P_Lu(p).
\end{equation}
The spatial terms $\gamma^i k_i$ vanish after angular integration,
$\int d\Omega\,\mathbf{k}=0$, while
$\gamma^0u(p)=u(p)$ in the rest frame. Since
$I_0=\bar{u}(q)P_Lu(p)$, the absorptive interference factor is therefore
\begin{equation}
2\,
\frac{
\operatorname{Im}(I_0^*I_1)
}{
|I_0|^2
}
=
\frac{1}{4\pi}
\frac{|\mathbf{k}|_\star}{M_i}
\left(E_j+M_j\right).
\label{eq:appendix_absorptive_factor}
\end{equation}

Substituting the on-shell expressions for $|\mathbf{k}|_\star$ and $E_j$
into Eq.~\eqref{eq:appendix_absorptive_factor} gives
\begin{align}
\frac{|\mathbf{k}|_\star}{M_i}
\left(E_j+M_j\right)
&=
\frac{\sqrt{\lambda(M_i^2,M_j^2,M_\phi^2)}}{2M_i^2}
\left(
\frac{M_i^2+M_j^2-M_\phi^2}{2M_i}
+
M_j
\right)
\nonumber\\
&\equiv
f(M_i^2,M_j^2,M_\phi^2).
\label{eq:appendix_f_relation}
\end{align}
It is therefore convenient to define
\begin{equation}
f(x,y,z)
=
\frac{\sqrt{\lambda(x,y,z)}}{2x}
\left(
\frac{x+y-z}{2\sqrt{x}}
+
\sqrt{y}
\right),
\label{eq:scalar_loop_function_appendix}
\end{equation}
where
\begin{equation}
\lambda(x,y,z)
=
x^2+y^2+z^2-2xy-2xz-2yz
\end{equation}
is the K\"all\'en function. Equation~\eqref{eq:appendix_absorptive_factor}
can thus be written compactly as
\begin{equation}
2\,
\frac{
\operatorname{Im}(I_0^*I_1)
}{
|I_0|^2
}
=
\frac{1}{4\pi}
f(M_i^2,M_j^2,M_\phi^2).
\label{eq:appendix_absorptive_f}
\end{equation}

For fixed internal states $N_j$ and $N_k$, the CP-odd coupling factor is
\begin{align}
\operatorname{Im}
\left(
\lambda_{0,\alpha}^{*}\lambda_{1,\alpha}
\right)
&=
-
\frac{M_i+M_k}{M_i^2-M_k^2}
\operatorname{Im}
\left[
Y_{\alpha i}^{*}Y_{\alpha k}
y_{ij}y_{jk}
\right].
\label{eq:appendix_coupling_im}
\end{align}
Substituting Eqs.~\eqref{eq:appendix_absorptive_f} and
\eqref{eq:appendix_coupling_im} into the interference expression
\eqref{eq:appendix_interference} gives, for a fixed pair $(j,k)$,
\begin{equation}
\epsilon_{i\alpha}^{\phi}(j,k)
=
\frac{1}{4\pi(Y^\dagger Y)_{ii}}
\operatorname{Im}
\left[
Y_{\alpha i}^{*}Y_{\alpha k}
y_{ij}y_{jk}
\right]
\frac{M_i+M_k}{M_i^2-M_k^2}
f(M_i^2,M_j^2,M_\phi^2).
\end{equation}
The absorptive cut requires $M_i>M_j+M_\phi$, so for the ordered spectrum
only states with $j<i$ contribute. Summing over all allowed intermediate
$N_j$ and virtual $N_k$ states therefore gives
\begin{equation}
\epsilon_{i\alpha}^{\phi}
=
\frac{1}{4\pi(Y^\dagger Y)_{ii}}
\sum_{\substack{j<i\\k\neq i}}
\operatorname{Im}
\left[
Y_{\alpha i}^{*}Y_{\alpha k}
y_{ij}y_{jk}
\right]
\frac{M_i+M_k}{M_i^2-M_k^2}
f(M_i^2,M_j^2,M_\phi^2).
\label{eq:epsilon_flavoured_appendix}
\end{equation}
The corresponding unflavoured asymmetry follows by summing over the
charged-lepton flavours. Using
\begin{equation}
\sum_\alpha
Y_{\alpha i}^{*}Y_{\alpha k}
=
(Y^\dagger Y)_{ik},
\end{equation}
we obtain
\begin{equation}
\epsilon_i^\phi
=
\frac{1}{4\pi(Y^\dagger Y)_{ii}}
\sum_{\substack{j<i\\k\neq i}}
\operatorname{Im}
\left[
(Y^\dagger Y)_{ik}
y_{ij}y_{jk}
\right]
\frac{M_i+M_k}{M_i^2-M_k^2}
f(M_i^2,M_j^2,M_\phi^2).
\label{eq:epsilon_unflavoured_appendix}
\end{equation}
We now derive an analytic upper bound on the $N_2$ contribution
in the hierarchical limit.
\subsection{Bound on $\epsilon_2^\phi$}
In the limit $M_i\gg M_j,M_\phi,$ the loop function in
Eq.~\eqref{eq:scalar_loop_function_appendix} reduces to
\begin{equation}
    f(M_i^2,M_j^2,M_\phi^2)
    \simeq
    \frac{M_i}{4}.
\end{equation}
The unflavoured scalar-induced asymmetry becomes
\begin{equation}
\epsilon_i^\phi
\simeq
\frac{M_i}
{16\pi(Y^\dagger Y)_{ii}}
\sum_{\substack{j<i\\k\neq i}}
\frac{
\operatorname{Im}
\left[
(Y^\dagger Y)_{ik}
y_{ij}y_{jk}
\right]
}{
M_i-M_k
}.
\label{eq:epsilon_hierarchical_appendix}
\end{equation}

Specialising to $i=2$, taking the scalar couplings $y_{ij}$ to be real,
and imposing $\tilde m_1=0$, the terms involving the first column of the
neutrino Yukawa matrix vanish. The surviving contribution corresponds to
$j=1$ and $k=3$, giving
\begin{equation}
\epsilon_2^\phi
=
\frac{M_2}
{16\pi(Y^\dagger Y)_{22}}
\frac{
y_{21}y_{13}\,
\operatorname{Im}(Y^\dagger Y)_{23}
}{
M_2-M_3
}.
\end{equation}
For a hierarchical spectrum $M_3\gg M_2$, this gives
\begin{equation}
|\epsilon_2^\phi|
\simeq
\frac{M_2}{16\pi M_3}
\frac{|y_{21}y_{13}|}
{(Y^\dagger Y)_{22}}
\left|
\operatorname{Im}(Y^\dagger Y)_{23}
\right|.
\end{equation}

Using the Casas--Ibarra parametrisation,
\begin{equation}
(Y^\dagger Y)_{ii}
=
\frac{M_i}{v^2}\tilde m_i,
\end{equation}
together with the Cauchy--Schwarz inequality,
\begin{align}
\left|
\operatorname{Im}(Y^\dagger Y)_{23}
\right|
&\leq
\left|
(Y^\dagger Y)_{23}
\right|
\nonumber\\
&\leq
\sqrt{
(Y^\dagger Y)_{22}
(Y^\dagger Y)_{33}
}
\nonumber\\
&=
\frac{\sqrt{M_2M_3}}{v^2}
\sqrt{\tilde m_2\tilde m_3},
\end{align}
we obtain
\begin{equation}
|\epsilon_2^\phi|_{\max}
=
\frac{|y_{21}y_{13}|}{16\pi}
\sqrt{\frac{M_2}{M_3}}
\sqrt{\frac{\tilde m_3}{\tilde m_2}}\ .
\label{eq:epsilon2_bound_appendix}
\end{equation}
The corresponding bound for an individual charged-lepton flavour follows
from
\begin{equation}
\left|
Y_{\alpha2}^{*}Y_{\alpha3}
\right|
\leq
|Y_{\alpha2}||Y_{\alpha3}|,
\end{equation}
and the definition
\begin{equation}
\tilde m_{i\alpha}
=
\frac{v^2|Y_{\alpha i}|^2}{M_i}.
\end{equation}
This gives
\begin{equation}
|\epsilon_{2\alpha}^{\phi}|_{\max}
=
\frac{|y_{21}y_{13}|}{16\pi}
\sqrt{\frac{M_2}{M_3}}
\frac{
\sqrt{\tilde m_{2\alpha}\tilde m_{3\alpha}}
}{
\tilde m_2
}\ .
\label{eq:epsilon2_flavoured_bound_appendix}
\end{equation}
The flavoured bounds are correlated through the common Yukawa matrix and
therefore cannot, in general, be saturated independently.

\subsection{CP-Asymmetry Density Matrix
$\epsilon_{\alpha\beta}^{(2),\phi}$}

For the density-matrix treatment, the flavour index of the lepton produced
in the decay amplitude must not be contracted with that of its conjugate.
Instead, the decay rates are promoted to matrices in flavour space. We
define
\begin{equation}
\Gamma_{\alpha\beta}
\propto
\mathcal{M}_\alpha\mathcal{M}_\beta^*,
\qquad
\overline{\Gamma}_{\alpha\beta}
\propto
\overline{\mathcal{M}}_\alpha^*
\overline{\mathcal{M}}_\beta ,
\end{equation}
so that the tree-level decay matrix has the same flavour-index convention
as the projector
$P_{\alpha\beta}^{0(i)}
=Y_{\alpha i}Y_{\beta i}^*/(Y^\dagger Y)_{ii}$.
The corresponding CP-asymmetry matrix is \cite{blanchet2013leptogenesisheavyneutrinoflavours}
\begin{equation}
\epsilon_{\alpha\beta}^{(i)}
=
\frac{
\Gamma_{\alpha\beta}
-
\overline{\Gamma}_{\alpha\beta}
}{
\displaystyle
\sum_\gamma
\left(
\Gamma_{\gamma\gamma}
+
\overline{\Gamma}_{\gamma\gamma}
\right)
}.
\label{eq:appendix_density_definition}
\end{equation}

Using
\begin{equation}
\mathcal{M}_\alpha
=
\lambda_{0,\alpha}I_0
+
\lambda_{1,\alpha}I_1,
\end{equation}
and retaining only the interference between the tree-level and one-loop
amplitudes gives
\begin{align}
\mathcal{M}_\alpha\mathcal{M}_\beta^*
={}&
\lambda_{0,\alpha}\lambda_{0,\beta}^*|I_0|^2
+
\lambda_{0,\alpha}\lambda_{1,\beta}^*I_0I_1^*
\nonumber\\
&+
\lambda_{1,\alpha}\lambda_{0,\beta}^*I_1I_0^*
+
\mathcal{O}(|I_1|^2).
\end{align}
For the CP-conjugate process the weak couplings are complex conjugated,
while the absorptive part of the loop is unchanged. The tree-level terms
cancel in the difference, yielding
\begin{equation}
\Gamma_{\alpha\beta}
-
\overline{\Gamma}_{\alpha\beta}
\propto
4\,\mathcal{J}_{\alpha\beta}
\,\operatorname{Im}(I_0^*I_1),
\end{equation}
where
\begin{equation}
\mathcal{J}_{\alpha\beta}
\equiv
\frac{1}{2i}
\left(
\lambda_{0,\alpha}\lambda_{1,\beta}^*
-
\lambda_{1,\alpha}\lambda_{0,\beta}^*
\right).
\label{eq:appendix_Jab}
\end{equation}
This quantity is Hermitian,
$\mathcal{J}_{\beta\alpha}^*=\mathcal{J}_{\alpha\beta}$.
For $\alpha=\beta$,
\begin{equation}
\mathcal{J}_{\alpha\alpha}
=
-\operatorname{Im}
\left(
\lambda_{0,\alpha}^*\lambda_{1,\alpha}
\right),
\end{equation}
so that the diagonal entries reproduce the flavoured interference
expression derived above.

At tree level, the denominator of
Eq.~\eqref{eq:appendix_density_definition} is proportional to
$2(Y^\dagger Y)_{ii}|I_0|^2$. Hence, for a fixed pair of internal
states $(j,k)$,
\begin{equation}
\epsilon_{\alpha\beta}^{(i),\phi}(j,k)
=
2\,
\frac{\mathcal{J}_{\alpha\beta}}
{(Y^\dagger Y)_{ii}}
\frac{
\operatorname{Im}(I_0^*I_1)
}{
|I_0|^2
}.
\label{eq:appendix_density_interference}
\end{equation}

For the scalar loop,
\begin{equation}
\lambda_{0,\alpha}=Y_{\alpha i},
\qquad
\lambda_{1,\alpha}
=
-
y_{ij}y_{jk}Y_{\alpha k}
\frac{M_i+M_k}{M_i^2-M_k^2}.
\end{equation}
Substituting these expressions into
Eq.~\eqref{eq:appendix_Jab} gives
\begin{align}
\mathcal{J}_{\alpha\beta}
={}&
\frac{M_i+M_k}{M_i^2-M_k^2}
\frac{1}{2i}
\Big[
Y_{\alpha k}Y_{\beta i}^*
\,y_{ij}y_{jk}- Y_{\alpha i}Y_{\beta k}^*
\,y_{ij}^*y_{jk}^*
\Big].
\label{eq:appendix_Jab_scalar}
\end{align}
Using the absorptive factor derived previously,
\begin{equation}
2\,
\frac{
\operatorname{Im}(I_0^*I_1)
}{
|I_0|^2
}
=
\frac{1}{4\pi}
f(M_i^2,M_j^2,M_\phi^2),
\end{equation}
we obtain
\begin{align}
\epsilon_{\alpha\beta}^{(i),\phi}
={}&
\frac{1}{4\pi(Y^\dagger Y)_{ii}}
\sum_{\substack{j<i\\k\neq i}}
\frac{M_i+M_k}{M_i^2-M_k^2}
f(M_i^2,M_j^2,M_\phi^2)
\nonumber\\
&\times
\frac{1}{2i}
\left[
Y_{\alpha k}Y_{\beta i}^*
\,y_{ij}y_{jk}
-
Y_{\alpha i}Y_{\beta k}^*
\,y_{ij}^*y_{jk}^*
\right].
\label{eq:epsilon_density_appendix}
\end{align}
The restriction $j<i$ follows from the absorptive threshold
$M_i>M_j+M_\phi$. Equation~\eqref{eq:epsilon_density_appendix} is
explicitly Hermitian.

For the $N_2$-dominated scenario considered in this work, the relevant
cut contains the intermediate $N_1$, while the virtual heavy-neutrino
state is $N_3$. Setting $i=2$, $j=1$, and $k=3$ gives
\begin{align}
\epsilon_{\alpha\beta}^{(2),\phi}
={}&
\frac{1}{4\pi(Y^\dagger Y)_{22}}
\frac{M_2+M_3}{M_2^2-M_3^2}
f(M_2^2,M_1^2,M_\phi^2)
\nonumber\\
&\times
\frac{1}{2i}
\left[
Y_{\alpha3}Y_{\beta2}^*\,y_{21}y_{13}
-
Y_{\alpha2}Y_{\beta3}^*\,y_{21}^*y_{13}^*
\right].
\label{eq:epsilon_density_N2_appendix}
\end{align}

For the real scalar couplings considered throughout this work and in the
hierarchical limit $M_3\gg M_2\gg M_1,M_\phi$, we use
$f(M_2^2,M_1^2,M_\phi^2)\simeq M_2/4$, giving
\begin{equation}
\epsilon_{\alpha\beta}^{(2),\phi}
\simeq
\frac{M_2}{16\pi(Y^\dagger Y)_{22}}
\frac{y_{21}y_{13}}{M_2-M_3}
\frac{1}{2i}
\left(
Y_{\alpha3}Y_{\beta2}^*
-
Y_{\alpha2}Y_{\beta3}^*
\right).
\label{eq:epsilon_density_hierarchical_appendix}
\end{equation}
As a consistency check, setting $\alpha=\beta$ gives
\begin{equation}
\frac{1}{2i}
\left(
Y_{\alpha3}Y_{\alpha2}^*
-
Y_{\alpha2}Y_{\alpha3}^*
\right)
=
\operatorname{Im}
\left(
Y_{\alpha2}^*Y_{\alpha3}
\right),
\end{equation}
and therefore
\begin{equation}
\epsilon_{\alpha\alpha}^{(2),\phi}
=
\epsilon_{2\alpha}^{\phi}.
\end{equation}
Similarly,
\begin{equation}
\sum_\alpha
Y_{\alpha2}^*Y_{\alpha3}
=
(Y^\dagger Y)_{23},
\end{equation}
so taking the trace gives
\begin{equation}
\operatorname{Tr}
\left[
\epsilon^{(2),\phi}
\right]
=
\epsilon_2^\phi.
\end{equation}
The density-matrix expression therefore contains the flavoured and
unflavoured scalar-induced CP asymmetries as its diagonal components and
trace, respectively, while its off-diagonal entries retain the flavour
coherence generated by the same tree--loop interference.
\bibliographystyle{unsrt}
\bibliography{references}

\end{document}